# Genomic and pathological analyses of an asymmetric true hermaphroditism case in a female labrador retriever

Yihang Zhou[1]

[1]*School of Life Sciences and Technology, Tongji University, Shanghai, China*

**Abstract**

The two main gonadal development disorders in dogs are true hermaphroditism and XX male syndrome. True hermaphroditism can be divided into two subcategories: XX sex reversal and XY sex reversal. XX Sry-negative sex reversal is more common, and it is characterized by the presence of both ovarian and testicular tissues in an animal. To date, there are 16 cases of true hermaphroditism reported in the literature, 15 of which are XX true hermaphroditism. Hermaphroditism has not been formally documented in labrador retrievers, and no case of asymmetric hermaphroditism has been reported in the literature.

**Introduction**

Canine gonad development is a fast and complex process that has many factors associated with it. During the development of a puppy in the uterus, there are three stages: ovum, embryo, and fetus. The time frame for a fertilized ovum to develop into a fetus is only about 61 days from start to finish. Typically, sex cannot be determined until day 30. Embryonic development is the period during development that the three germ layers differentiate into different tissues. The three germ layers, ectoderm, mesoderm, and endoderm, all develop to form critical parts of the animal's body. This process of embryologic development dividing into three layers is called gastrulation. Mesoderm forms most of the reproductive and urogenital tract such as the gonads, uterus, cervix, part of the vagina, epididymis, and accessory sex glands. The ectoderm forms another part of the reproductive tract, such as the mammary glands, hypothalamus, pituitary, vestibule, penis, clitoris, and part of the vagina. At the beginning of sex differentiation,

The female reproductive tract is the base template for sex differentiation--without any hormone or other intervention, a female reproductive tract will develop naturally. When there is no Y chromosome, SRY gene, TDF factor, or AMH, the Wolffian duct is not supported and degenerates. The Mullerian duct is supported and continues to develop into the female internal reproductive tract. The female gonads originate from primordial germ cells (PGCs) that migrate to the gonadal ridge from the mesoderm. The PCGs continue to differentiate into oocytes that make up the ovaries of the female. As long as Anti-Mullerian hormone (AMH) and Testis Determining Factor (TDF) are absent, the female reproductive tract will develop normally both externally and internally.

The male reproductive tract must be initiated by hormones to develop properly. When a fertilized egg's chromosomes are XY instead of XX, the chromosomal designation is male

When the chromosomal designation is XY, or male, there is a specific cascade of events that occurs so that the proper internal and external reproductive tract forms. The SRY gene is located on the Y chromosome, and releases TDF. TDF then kickstarts the gonads to develop as testes instead of ovaries, and the animal will be genotypically and phenotypically a male. (TDF) helps to differentiate the indifferent gonad, a precursor to both male and female gonads that is still bipotential. Once the testes start to form in the embryo, they begin secreting hormones. The first hormone produced and secreted by these testes is AMH, also called Mullerian inhibiting hormone (MIH). This hormone acts on the Mullerian cords and causes them to regress so that the wolffian cords can develop and form the male internal duct. When the Mullerian cords regress, this is ensuring that the animal does not develop a female internal tract.  Another hormone the testes produce is testosterone (T). Testosterone is produced from the Leydig cells, so those cells must be formed by bipotential cells that differentiate into Leydig cells. Testosterone causes the formation of epididymis and ductus deferens, and DHT, a form of testosterone, triggers the formation of the urethra, prostate, penis, prepuce, and scrotum. Since the male reproductive system is not the base template of reproductive development, it must be encouraged and catalyzed by male hormones to ensure normal development.

There are three categories of reproductive abnormalities: chromosomal disorders, gonadal development disorders, and phenotypical development disorders. Chromosomal disorders result from issues with the number or structure of the animal's sex chromosome makeup. These disorders include Monosomy X, Trisomy X, Klinefelter's syndrome, XX/XY chimera and mosaics, and XX/XO. Monosomy X, also called Turner's syndrome in humans, is very uncommon in dogs. Monosomy X typically presents as normal phenotypic females with infertility issues and stunted growth. Trisomy X, or XXX, also present typical phenotypic female signs, but have varying stages of ovarian development. Most dogs with Trisomy X also do not have typical estrus cycles, have hypoplastic ovaries, or no normal follicles on the ovaries. There are many different types of XX/XY mosaics and chimeras, which are also quite uncommon.

Gonadal development disorders are typically more common than chromosomal disorders, or at least better documented due to these disorders being more phenotypically obvious. The two main gonadal development disorders are XX true hermaphroditism and XX male syndrome.

XX true hermaphroditism, also known as XX Sex Reversal, can be divided into two subcategories—XX sex reversal and XY sex reversal. XX sex reversal is much more common, and XY rex reversal is quite rare, with only a few cases reported. XX sex reversal, or XX true hermaphroditism, is characterized by both ovarian and testicular tissue in the animal. This combination of gonadal tissues can be expressed bilaterally, unilaterally, and laterally. This is not to be confused with pseudohermaphroditism, which is when an animal has a certain phenotype, and the opposite sex gonadal tissues.

Phenotypical development disorders are the third class of sex disorders, and their chromosomal and gonadal sex match, but the animal's phenotype is expressed incorrectly for their sex. An example of a phenotypical development disorder is pseudohermaphroditism. Pseudohermaphroditism occurs when an animal has a chromosomal and gonadal sex that match each other, but the external genital tract and characteristics are of the opposite sex. There can be male and female pseudohermaphroditism, but male pseudohermaphroditism occurs more frequently. Pseudohermaphroditism is often caused by hormones being produced and expressed incorrectly, for example, Persistent Mullerian Duct Syndrome (PMDS) or Androgen Insensitivity Syndrome.

primordial germ cells (PGCs) are created from the mesoderm and eventually form sperm or eggs depending on the animal. Primordial germ cells have three phases--development, increasing PGC numbers, and sex differentiation. This literature review will explore male and female gonad development in dogs, as well as various canine reproductive abnormalities.

A six-week-old hermaphrodite labrador retriever born in 2006 was admitted with a history of "two tails", and stool exsits from a vestibule. This rare congenital abnormality was further examined, and two rounds of surgeries corrected the genitalia, bladder, and gut abnormalities. Karyotyping and sex-specific PCR assay were performed using blood samples. After she passed away in 2016, necropsy and histopathological analyses were performed. Genomic DNA was extracted from FFPE samples from male-specific prepuce tissues, female-specific vulva, and a combination of 4 somatic tissues. Whole-genome resequencing (WGR) and de novo SNP calling were performed on these three types of tissues to identify potential genetic basis of this unique hermaphroditism.

Karyotyping analysis confirmed 78, XX (Sry-), which is normal female without any apparent chromosomal abnormalities. Physical examination of the puppy found an imperforate anus, male/female extrenal genitalia, and perianal asymmetry. A rudimentary scrotum (the second "tail") with a small prepuce and penis were located next to a vulva and vestibule. Abdomen radiology identified a rectovaginal fistula and the caudal rectum appeared to extend to the membrane covering the anal sphincter, which is Type 2 Atresia ani with an overall incidence of 70 in 1,000,000 dogs. Excretory urogram identified a right ectopic ureter and right and left urinary bladders separated by a fibromuscular septum. The sagittal duplication of bladders was documented in human with only 50 cases reported. A ureteroneocystostomy was performed to correct this. The female-side of the reproductive system consists of two normal ovaries, separate uterine horns and cervis connected to the vagina. The male-side includes a rudimentary scrotum, penis, and prepuce. As an adult, the dog continued to urinate from both the penis and the vulva. >99% of WGR reads were aligned to the dog reference genome, and achieved 27.5× genome coverage of male-specific tissue, 22.2× in somatic tissues, and 46.3 in female-specific tissue. De novo SNP calling identified 1.7 million SNPs compared to the referenc genome, of which only 0.15% genotypes differ between male and female/somatic tissues. Further genotyping and SNP consequence analyses confirmed that these are mainly false positives, and targeted analysis of dog sex-reversal genes did not identify any candidate changes. Our results indicate that this rare asymmetric hermaphroditism is unlikely to be genetic, and it is presumably due to errors in the early developmental reprogramming.

Normal dogs have 78 chromosomes, including the sex chromosomes. The 78,XX is female, and 78,XY is male. Chromosomal abnormalities will cause disorders of sexual development (DSD), which includes Turner syndrome (or Monosomy X 77,XO) [1], Trisomy X (79, XXX) [2], and Klinefelter syndrome (79,XXY) [3]. Chimerism and mosaicism also cause DSD [4]. DSD has been reported in dogs in three categories, the sex chromosome DSD, the XY DSD, and the XX DSD [5, 6] .

DSD is recognized as a complex procedure, nevertheless, we are not clear about the mechanism [7-9]. Sexually dimorphic development is a multi-factor procedure [10]. The repressing factors and activating factors should interacte precisely in spatio-temporal patterns to ensure normal sex development [11]. The Sex-determining Region Y (SRY in humans, Sry in mammals) gene switches binarily as the initiation in male sex development [12]. In males, SRY induces SRY-box transcription factor 9 (SOX9) to trigger the bipotential gonad and maintain the differentiation pathway to testis [13]. In females, the Forkhead transcription factor 2 (FOXL2), the R-spondin-1 (RSPO1), and the Wingless type MMTV integration site family member 4 (WNT4), together, initiate the female sexual development [14]. Other sexual development factors include, but not limited to Wilms' tumor 1 (WT1), LIM homeobox factor 9 (LHX9), GATA binding protein 4 (GATA4), empty spiracles homeobox2 (EMX2), and chromobox homolog2 (CBX2) [15]. Any disorders in the complex regulatory network may lead to unexpected impact in sex development [16].

Despite the many reports of DSD in dogs, our knowledge of DSD at the molecular level is mainly from human studies. For these dog cases, the veterinarian and pathologist usually will do the physical review, genital examination, histological check, cytogenetic analysis for karyotyping, and polymerase chain reaction (PCR) for target known genes (such as SRY and SOX9) [17]. Though these tests are standard and sufficient to cure or report cases, they are not enough to expand our knowledge to the unknown region. What mutations exist in a DSD dog's genome? Are there unknown genetic components evolved in DSD dogs? Does epigenetics factor have an impact on the disorder?

To address these concerns, it is urgent to apply newer technologies in DSD dog research. In this article, we applied whole-genome sequencing to a DSD dog named Xugo and executed a comprehensive SNPs analysis. To our knowledge, this work is the first to offer a whole genome map for a DSD dog. Our analysis results shall contribute to the DSD research community.

**Materials and Methods**

**Case**

Xugo is a female labrador from a healthy litter of seven.

**Karyotyping**

Xugo's Blood sample was sent for karyotyping analysis in July 2008. The chromosome number and morphology were analyzed using Giemsa staining in while blood cells. PCR amplification was performed using primers targeting the *SRY* gene on the Y chromosome and the X-linked *AR* gene (androgen receptor), with male control, female control, and negative control templates.

**Genomic DNA extractions from FFPE samples**

Slides were cut from formalin-fixed paraffin-embedded (FFPE) samples of kidney Genomic DNA (gDNA) was extracted from these tissues with a total surface area ~20 mm$^2$ using Zymo Quick-DNA/RNA FFPE Miniprep kit (Zymo Research, USA) following the manufacture protocols. The yield and quality of extracted gDNA were checked with NanoDrop and Qubit 3.0 Fluorometer (Thermo Fisher Scientific, USA).

**Genome resequencing library preparation and sequencing**

The gDNA resequencing library were constructed was used in Illumina paired end short-read DNA library preparation (target 300 bp insert size) with the NEBNext Ultra II FS DNA Library Prep Kit (NEB # E7805, #E6177). After we completed the library construction, the concentration was measured with Qubit 3.0 Fluorometer and the size distribution was assessed using Agilent TapeStation 4200 (Agilent technologies, CA). The prepared gDNA library was sequenced on a NovaSeq sequencer using Illumina NovaSeq S4 platform at the Genomic Services Lab at the HudsonAlpha Institute for Biotechnology.

**Upstream reads process**

We sequenced 7 samples, including 1 kidney sample, 1 rectocolic junction sample, 1 lung heart mix sample, 2 vulva samples and 2 preuce samples, which grouped into somatic, female and male specific tissues. The raw reads were trimmed using Trimmomatic (v 0.36) to remove adapters and low quality bases or reads with the minimum length threshold set to 32bp. The filtered reads were then mapped to CanFam 3.1 genome using bwa (v 0.7.17) to have sam files, samtools (v 1.6) was then applied to generate bam file. With the 7 bam files in hand, we merged them into somatic-bam, female-bam and male-bam.

**SNP calling and analyses**

Then we applied the SNP-calling pipeline of GATK (v 3.8) to generate vcf files and a bunch of intermediate files. Due to the fact that SNPs calling is always having false positive calls in practice, we used Intergrative Genome Viewer (IGV, v 2.8.9) and mannually check the calling results to determined the threshold of QUAL=500 and COV=6. Further more, we removed short aligned reads whose mapped region is less than 75bp to reduce noise caused by CIGAR softclip. Then we used samtools and self-written scripts to generate the SNPs summary table for following analysis.

For each SNP position, we labled "0" for homozygous reference genotype (refer allele/refer allele), "1" for heterozygous genotype (alternative allele/refer allele) and "2" for homozygous alternative genotype (alternative allele/alternative allele). We excluded SNPs whose reference or alternative alleles are inconsistent among samples, or with a too low coverage less than 6. The SNP summary table was then transformed into bed format to be annotated using Snpeff (v 4.3). Logically, the hermaphroditic phenotype mutation, if exists, should be female tissue, male tissue or somatic tissue specific. Therefore, we extracted missense SNPs and high impact SNPs of "122" (female-male-somatic), "212" and "221" from the annotation. We extracted 17 missense SNPs from "122", 46 missense SNPs from "212" and 27 missense SNPs. These SNPs were processed using KEGG, but no significant pathways were enriched. Only one high impact SNP was called, which is the mutation of STS located at chromosome X, 4466440. We then mannually checked the SNPs in IGV, but it turned out to be a false positive case.

Despite that we could not identify SNPs causing the hermaphroditic phenotype, this result itself strongly indicates that this phenotype is not the result of any SNPs.

**Results**

**An extremely rare case of asymmetric hermaphroditism in a female labrador.**

A six-week-old Chocolate Labrador puppy was admitted with a history of "two-tails". Physical examination of the puppy found an imperforate anus, male and female external genitalia, and perianal asymmetry (the anus was slightly right of midline with a pouch of soft tissue skin to the left). A rudimentary scrotum (the second "tail") with a small prepuce and penis were located next to a vulva and vestibule. An unhaired line of skin was distinct from the ventral anus to the dorsal vulva. A rectovaginal fistula was suspected as feces were noted within the vulva. Otherwise, the examination of the puppy was normal.

**Karyotyping results confirm normal female chromosomal configuration (78, XX)**

The dog lived to be 11-years-old when it died of acute necrotizing hemorrhagic pancreatitis.

**Type 2 Atresia ani and surgical correction**

Radiographs of the abdomen were obtained after injecting contrast into the vestibule, the rectovaginal fistula was identified, and the caudal rectum appeared to extend to the membrane covering the anal sphincter. Surgical correction of both the imperforate anus and the rectovaginal fistula was successful, and the puppy was fecal continent.

**Sagittal duplication of urinary bladders and surgical correction**

As the puppy grew, it was noted that urine puddles were left where the puppy sat or slept. During micturition, a urine stream was noted. Ureteral ectopia was suspected. It was also noted that urine exited the penis as well when posturing to urinate. At seven months of age, an excretory urogram was performed that identified a right ectopic ureter and right and left urinary bladders separated by a septum.

**Discussion**

This case describes an extremely rare example of asymmetric hermaphroditism in a dog, featuring both male and female external genitalia, anal atresia, and bladder duplication. Karyotyping and PCR analyses confirmed a normal female 78,XX genotype with no SRY gene, while whole-genome sequencing revealed no obvious mutations in known sex-determination genes. These results indicate that the condition likely arose from epigenetic or embryologic processes rather than a straightforward genetic anomaly.

Early recognition was critical for successful surgical correction of anal atresia and urinary tract malformations, improving the dog's quality of life. The atypical morphology underscores the complex spatiotemporal interplay of sex-determining factors during canine embryonic development. Importantly, this study highlights how advanced genomic tests may fail to detect subtle regulatory or epigenetic disruptions.

Further investigation into histone modifications, DNA methylation, or non-coding RNAs may help clarify the underlying mechanisms. This singular case emphasizes the need for a multi-disciplinary approach—combining clinical observations, pathological assessment, cytogenetics, and next-generation sequencing—when evaluating rare DSD presentations. These findings not only inform clinical management but also enhance our understanding of canine sex development as a model for other mammalian species.

**Declaration of Conflicting Interests**

# Reference

1.      Kesler, S.R., *Turner syndrome.* Child and Adolescent Psychiatric Clinics of North America, 2007. **16**(3): p. 709-722.
2.      Tartaglia, N.R., et al., *A review of trisomy X (47, XXX).* Orphanet journal of rare diseases, 2010. **5**(1): p. 1-9.
3.      Lanfranco, F., et al., *Klinefelter's syndrome.* The Lancet, 2004. **364**(9430): p. 273-283.
4.      Malan, V., M. Vekemans, and C. Turleau, *Chimera and other fertilization errors.* Clinical genetics, 2006. **70**(5): p. 363-373.
5.      Poth, T., et al., *Disorders of sex development in the dog—adoption of a new nomenclature and reclassification of reported cases.* Animal Reproduction Science, 2010. **121**(3-4): p. 197-207.
6.      Meyers-Wallen, V., *Gonadal and sex differentiation abnormalities of dogs and cats.* Sexual Development, 2012. **6**(1-3): p. 46-60.
7.      Biason-Lauber, A., *Control of sex development.* Best Pract Res Clin Endocrinol Metab, 2010. **24**(2): p. 163-86.
8.      Chassot, A.A., et al., *Genetics of ovarian differentiation: Rspo1, a major player.* Sex Dev, 2008. **2**(4-5): p. 219-27.
9.      DeFalco, T. and B. Capel, *Gonad morphogenesis in vertebrates: divergent means to a convergent end.* Annu Rev Cell Dev Biol, 2009. **25**: p. 457-82.
10.     Witchel, S.F., *Disorders of sex development.* Best Practice & Research Clinical Obstetrics & Gynaecology, 2018. **48**: p. 90-102.
11.     Biason-Lauber, A., *Control of sex development.* Best practice & research Clinical endocrinology & metabolism, 2010. **24**(2): p. 163-186.
12.     Koopman, P., et al., *Male development of chromosomally female mice transgenic for Sry.* Nature, 1991. **351**(6322): p. 117-121.
13.     Kanai, Y., et al., *From SRY to SOX9: Mammalian testis differentiation.* Journal of Biochemistry, 2005. **138**(1): p. 13-19.
14.     Biason-Lauber, A. *WNT4, RSPO1, and FOXL2 in sex development*. in *Seminars in reproductive medicine*. 2012. Thieme Medical Publishers.
15.     Rudigier, L.J., et al., *Ex vivo cultures combined with vivo-morpholino induced gene knockdown provide a system to assess the role of WT1 and GATA4 during gonad differentiation.* PloS one, 2017. **12**(4): p. e0176296.
16.     Lin, Y.-T. and B. Capel, *Cell fate commitment during mammalian sex determination.* Current Opinion in Genetics & Development, 2015. **32**: p. 144-152.
17.     Szczerbal, I., et al., *Chromosome abnormalities in dogs with disorders of sex development (DSD).* Animal reproduction science, 2021. **230**: p. 106771.

1.      Kesler, S.R., *Turner syndrome.* Child and Adolescent Psychiatric Clinics of North America, 2007. **16**(3): p. 709-722.
2.      Tartaglia, N.R., et al., *A review of trisomy X (47, XXX).* Orphanet journal of rare diseases, 2010. **5**(1): p. 1-9.
3.      Lanfranco, F., et al., *Klinefelter's syndrome.* The Lancet, 2004. **364**(9430): p. 273-283.
4.      Malan, V., M. Vekemans, and C. Turleau, *Chimera and other fertilization errors.* Clinical genetics, 2006. **70**(5): p. 363-373.
5.      Poth, T., et al., *Disorders of sex development in the dog—adoption of a new nomenclature and reclassification of reported cases.* Animal Reproduction Science, 2010. **121**(3-4): p. 197-207.
6.      Meyers-Wallen, V., *Gonadal and sex differentiation abnormalities of dogs and cats.* Sexual Development, 2012. **6**(1-3): p. 46-60.

7.      Biason-Lauber, A., *Control of sex development.* Best Pract Res Clin Endocrinol Metab, 2010. **24**(2): p. 163-86.

8.      Chassot, A.A., et al., *Genetics of ovarian differentiation: Rspo1, a major player.* Sex Dev, 2008. **2**(4-5): p. 219-27.

9.      DeFalco, T. and B. Capel, *Gonad morphogenesis in vertebrates: divergent means to a convergent end.* Annu Rev Cell Dev Biol, 2009. **25**: p. 457-82.

10.     Witchel, S.F., *Disorders of sex development.* Best Practice & Research Clinical Obstetrics & Gynaecology, 2018. **48**: p. 90-102.

11.     Biason-Lauber, A., *Control of sex development.* Best practice & research Clinical endocrinology & metabolism, 2010. **24**(2): p. 163-186.

12.     Koopman, P., et al., *Male development of chromosomally female mice transgenic for Sry.* Nature, 1991. **351**(6322): p. 117-121.

13.     Kanai, Y., et al., *From SRY to SOX9: Mammalian testis differentiation.* Journal of Biochemistry, 2005. **138**(1): p. 13-19.

14.     Biason-Lauber, A. *WNT4, RSPO1, and FOXL2 in sex development*. in *Seminars in reproductive medicine*. 2012. Thieme Medical Publishers.

15.     Rudigier, L.J., et al., *Ex vivo cultures combined with vivo-morpholino induced gene knockdown provide a system to assess the role of WT1 and GATA4 during gonad differentiation.* PloS one, 2017. **12**(4): p. e0176296.

16.     Lin, Y.-T. and B. Capel, *Cell fate commitment during mammalian sex determination.* Current Opinion in Genetics & Development, 2015. **32**: p. 144-152.

1.      Kesler, S.R., *Turner syndrome.* Child and Adolescent Psychiatric Clinics of North America, 2007. **16**(3): p. 709-722.

2.      Tartaglia, N.R., et al., *A review of trisomy X (47, XXX).* Orphanet journal of rare diseases, 2010. **5**(1): p. 1-9.

3.      Lanfranco, F., et al., *Klinefelter's syndrome.* The Lancet, 2004. **364**(9430): p. 273-283.

4.      Malan, V., M. Vekemans, and C. Turleau, *Chimera and other fertilization errors.* Clinical genetics, 2006. **70**(5): p. 363-373.

5.      Poth, T., et al., *Disorders of sex development in the dog—adoption of a new nomenclature and reclassification of reported cases.* Animal Reproduction Science, 2010. **121**(3-4): p. 197-207.

6.      Meyers-Wallen, V., *Gonadal and sex differentiation abnormalities of dogs and cats.* Sexual Development, 2012. **6**(1-3): p. 46-60.

7.      Biason-Lauber, A., *Control of sex development.* Best Pract Res Clin Endocrinol Metab, 2010. **24**(2): p. 163-86.

8.      Chassot, A.A., et al., *Genetics of ovarian differentiation: Rspo1, a major player.* Sex Dev, 2008. **2**(4-5): p. 219-27.

9.      DeFalco, T. and B. Capel, *Gonad morphogenesis in vertebrates: divergent means to a convergent end.* Annu Rev Cell Dev Biol, 2009. **25**: p. 457-82.

10.     Witchel, S.F., *Disorders of sex development.* Best Practice & Research Clinical Obstetrics & Gynaecology, 2018. **48**: p. 90-102.

11.     Biason-Lauber, A., *Control of sex development.* Best practice & research Clinical endocrinology & metabolism, 2010. **24**(2): p. 163-186.

12.     Koopman, P., et al., *Male development of chromosomally female mice transgenic for Sry.* Nature, 1991. **351**(6322): p. 117-121.

13.     Kanai, Y., et al., *From SRY to SOX9: Mammalian testis differentiation.* Journal of Biochemistry, 2005. **138**(1): p. 13-19.

14.    Biason-Lauber, A. *WNT4, RSPO1, and FOXL2 in sex development*. in *Seminars in reproductive medicine*. 2012. Thieme Medical Publishers.

1.     Kesler, S.R., *Turner syndrome.* Child and Adolescent Psychiatric Clinics of North America, 2007. **16**(3): p. 709-722.
2.     Tartaglia, N.R., et al., *A review of trisomy X (47, XXX).* Orphanet journal of rare diseases, 2010. **5**(1): p. 1-9.
3.     Lanfranco, F., et al., *Klinefelter's syndrome.* The Lancet, 2004. **364**(9430): p. 273-283.
4.     Malan, V., M. Vekemans, and C. Turleau, *Chimera and other fertilization errors.* Clinical genetics, 2006. **70**(5): p. 363-373.
5.     Poth, T., et al., *Disorders of sex development in the dog—adoption of a new nomenclature and reclassification of reported cases.* Animal Reproduction Science, 2010. **121**(3-4): p. 197-207.
6.     Meyers-Wallen, V., *Gonadal and sex differentiation abnormalities of dogs and cats.* Sexual Development, 2012. **6**(1-3): p. 46-60.
7.     Biason-Lauber, A., *Control of sex development.* Best Pract Res Clin Endocrinol Metab, 2010. **24**(2): p. 163-86.
8.     Chassot, A.A., et al., *Genetics of ovarian differentiation: Rspo1, a major player.* Sex Dev, 2008. **2**(4-5): p. 219-27.
9.     DeFalco, T. and B. Capel, *Gonad morphogenesis in vertebrates: divergent means to a convergent end.* Annu Rev Cell Dev Biol, 2009. **25**: p. 457-82.
10.    Witchel, S.F., *Disorders of sex development.* Best Practice & Research Clinical Obstetrics & Gynaecology, 2018. **48**: p. 90-102.
11.    Biason-Lauber, A., *Control of sex development.* Best practice & research Clinical endocrinology & metabolism, 2010. **24**(2): p. 163-186.
12.    Koopman, P., et al., *Male development of chromosomally female mice transgenic for Sry.* Nature, 1991. **351**(6322): p. 117-121.
13.    Kanai, Y., et al., *From SRY to SOX9: Mammalian testis differentiation.* Journal of Biochemistry, 2005. **138**(1): p. 13-19.

1.     Poth, T., et al., *Disorders of sex development in the dog—adoption of a new nomenclature and reclassification of reported cases.* Animal Reproduction Science, 2010. **121**(3-4): p. 197-207.
2.     Meyers-Wallen, V., *Gonadal and sex differentiation abnormalities of dogs and cats.* Sexual Development, 2012. **6**(1-3): p. 46-60.
3.     Biason-Lauber, A., *Control of sex development.* Best Pract Res Clin Endocrinol Metab, 2010. **24**(2): p. 163-86.
4.     Chassot, A.A., et al., *Genetics of ovarian differentiation: Rspo1, a major player.* Sex Dev, 2008. **2**(4-5): p. 219-27.
5.     DeFalco, T. and B. Capel, *Gonad morphogenesis in vertebrates: divergent means to a convergent end.* Annu Rev Cell Dev Biol, 2009. **25**: p. 457-82.
6.     Kesler, S.R., *Turner syndrome.* Child and Adolescent Psychiatric Clinics of North America, 2007. **16**(3): p. 709-722.
7.     Tartaglia, N.R., et al., *A review of trisomy X (47, XXX).* Orphanet journal of rare diseases, 2010. **5**(1): p. 1-9.
8.     Lanfranco, F., et al., *Klinefelter's syndrome.* The Lancet, 2004. **364**(9430): p. 273-283.
9.     Malan, V., M. Vekemans, and C. Turleau, *Chimera and other fertilization errors.* Clinical genetics, 2006. **70**(5): p. 363-373.

1.     Biason-Lauber, A., *Control of sex development.* Best Pract Res Clin Endocrinol Metab, 2010. **24**(2): p. 163-86.

2.      Chassot, A.A., et al., *Genetics of ovarian differentiation: Rspo1, a major player.* Sex Dev, 2008. **2**(4-5): p. 219-27.
3.      DeFalco, T. and B. Capel, *Gonad morphogenesis in vertebrates: divergent means to a convergent end.* Annu Rev Cell Dev Biol, 2009. **25**: p. 457-82.
4.      Kesler, S.R., *Turner syndrome.* Child and Adolescent Psychiatric Clinics of North America, 2007. **16**(3): p. 709-722.
5.      Tartaglia, N.R., et al., *A review of trisomy X (47, XXX).* Orphanet journal of rare diseases, 2010. **5**(1): p. 1-9.
6.      Lanfranco, F., et al., *Klinefelter's syndrome.* The Lancet, 2004. **364**(9430): p. 273-283.
7.      Malan, V., M. Vekemans, and C. Turleau, *Chimera and other fertilization errors.* Clinical genetics, 2006. **70**(5): p. 363-373.

1.      Biason-Lauber, A., *Control of sex development.* Best Pract Res Clin Endocrinol Metab, 2010. **24**(2): p. 163-86.
2.      Chassot, A.A., et al., *Genetics of ovarian differentiation: Rspo1, a major player.* Sex Dev, 2008. **2**(4-5): p. 219-27.
3.      DeFalco, T. and B. Capel, *Gonad morphogenesis in vertebrates: divergent means to a convergent end.* Annu Rev Cell Dev Biol, 2009. **25**: p. 457-82.
4.      Kesler, S.R., *Turner syndrome.* Child and Adolescent Psychiatric Clinics of North America, 2007. **16**(3): p. 709-722.
5.      Tartaglia, N.R., et al., *A review of trisomy X (47, XXX).* Orphanet journal of rare diseases, 2010. **5**(1): p. 1-9.
6.      Lanfranco, F., et al., *Klinefelter's syndrome.* The Lancet, 2004. **364**(9430): p. 273-283.

1.      Biason-Lauber, A., *Control of sex development.* Best Pract Res Clin Endocrinol Metab, 2010. **24**(2): p. 163-86.
2.      Chassot, A.A., et al., *Genetics of ovarian differentiation: Rspo1, a major player.* Sex Dev, 2008. **2**(4-5): p. 219-27.
3.      DeFalco, T. and B. Capel, *Gonad morphogenesis in vertebrates: divergent means to a convergent end.* Annu Rev Cell Dev Biol, 2009. **25**: p. 457-82.
4.      Kesler, S.R., *Turner syndrome.* Child and Adolescent Psychiatric Clinics of North America, 2007. **16**(3): p. 709-722.
5.      Tartaglia, N.R., et al., *A review of trisomy X (47, XXX).* Orphanet journal of rare diseases, 2010. **5**(1): p. 1-9.

1.      Biason-Lauber, A., *Control of sex development.* Best Pract Res Clin Endocrinol Metab, 2010. **24**(2): p. 163-86.
2.      Chassot, A.A., et al., *Genetics of ovarian differentiation: Rspo1, a major player.* Sex Dev, 2008. **2**(4-5): p. 219-27.
3.      DeFalco, T. and B. Capel, *Gonad morphogenesis in vertebrates: divergent means to a convergent end.* Annu Rev Cell Dev Biol, 2009. **25**: p. 457-82.
4.      Kesler, S.R., *Turner syndrome.* Child and Adolescent Psychiatric Clinics of North America, 2007. **16**(3): p. 709-722.

1.      Biason-Lauber, A., *Control of sex development.* Best Pract Res Clin Endocrinol Metab, 2010. **24**(2): p. 163-86.
2.      Chassot, A.A., et al., *Genetics of ovarian differentiation: Rspo1, a major player.* Sex Dev, 2008. **2**(4-5): p. 219-27.

3.      DeFalco, T. and B. Capel, *Gonad morphogenesis in vertebrates: divergent means to a convergent end.* Annu Rev Cell Dev Biol, 2009. **25**: p. 457-82.

1.      Biason-Lauber, A., *Control of sex development.* Best Pract Res Clin Endocrinol Metab, 2010. **24**(2): p. 163-86.
2.      Chassot, A.A., et al., *Genetics of ovarian differentiation: Rspo1, a major player.* Sex Dev, 2008. **2**(4-5): p. 219-27.

Biason-Lauber, A. (2010). Control of sex development. *Best Pract Res Clin Endocrinol Metab, 24*(2), 163-186. doi:10.1016/j.beem.2009.12.002

Chassot, A. A., Gregoire, E. P., Magliano, M., Lavery, R., & Chaboissier, M. C. (2008). Genetics of ovarian differentiation: Rspo1, a major player. *Sex Dev, 2*(4-5), 219-227. doi:10.1159/000152038